\documentclass[a4paper]{article}
\usepackage{ASVspoof}
\usepackage{epsfig,amssymb,amsmath}
\usepackage{booktabs}
\usepackage{hyperref}
\ninept

\title{Augmentation through Laundering Attacks for Audio Spoof Detection}

\makeatletter
\def\name#1{\gdef\@name{#1\\}}
\makeatother
\name{{\em Hashim Ali, Surya Subramani, and Hafiz Malik}}

\address{Department of Electrical Engineering  \\
University of Michigan-Dearborn, Dearborn \\
{\small \tt \{hashim, suryasss, hafiz\}@umich.edu} }
\begin{document}
\maketitle

\begin{abstract}
Recent text-to-speech (TTS) developments have made voice cloning (VC) more realistic, affordable, and easily accessible. This has given rise to many potential abuses of this technology, including Joe Biden's New Hampshire deepfake robocall. Several methodologies have been proposed to detect such clones. However, these methodologies have been trained and evaluated on relatively clean databases. Recently, ASVspoof 5 Challenge introduced a new crowd-sourced database of diverse acoustic conditions including various spoofing attacks and codec conditions. This paper is our submission to the ASVspoof 5 Challenge and aims to investigate the performance of Audio Spoof Detection, trained using data augmentation through laundering attacks, on the ASVSpoof 5 database. The results demonstrate that our system performs worst on A18, A19, A20, A26, and A30 spoofing attacks and in the codec and compression conditions of C08, C09, and C10.

\end{abstract}

\section{Introduction} \label{intro}
Recent developments in text-to-speech (TTS) technology, particularly zero-shot, multi-speaker TTS \cite{casanova2022yourtts, wang2301neural, li2023zse}, have led to the creation of methods that can generate highly realistic synthesized speech. This progress has spurred the growth of companies such as ElevenLabs that provide affordable and easy-to-use TTS services. These advances facilitate a wide range of applications, from helping people with speech impairments to creating digital avatars, as demonstrated by a jailed Pakistani politician, Imran Khan, who created his AI-generated video for his election campaign \cite{ray_imran_nodate}. However, alongside the positive use cases, the potential for misuse of voice cloning (VC) technology has also raised many concerns.

The past two years have seen a remarkable increase in TTS/VC incidents targeting political figures. Recently, around 25000 voters in New Hampshire received a deepfake robocall impersonating President Joe Biden, telling them not to vote in the state's primary elections. This robocall was analyzed by a security company, called Pindrop, and it was attributed to be likely generated through Elevenlabs' technology \cite{ knibbs_researchers_nodate, balasubramaniyan_pindrop_2024, elliott_biden_nodate}. This kind of deepfake content is not just spread by hidden bad actors, it is also shared by many renowned people. For example, Elon Musk shared a sarcastic ``campaign video'' of Vice President Kamala Harris, in which she made comments along the lines of ``The first rule President Joe Biden taught me is to carefully hide your total incompetence'' and ``I believe exploring the significance of the insignificance is in itself significant'' \cite{Musk_X_Kamala, apnewsParodyShared}. Similarly, audio deepfakes of Donald Trump have also been shared on social networks \cite{Donald_parrot}. Other similar instances of targeting political figures include Ukrainian President Zelenskyy's viral deepfake video asking his soldiers to surrender \cite{ebaker_russian_2022} and mayor of London UK, Sadiq Khan's fake audio \cite{noauthor_sadiq_2024} in which he was supposedly making inflammatory remarks about Armistice Day and rallying people to protest for Palestine. In addition, audio deepfakes are being used in phone scams in which a person receives a call from a scammer claiming to be a relative stuck in an accident, arrest, or abduction to extort money from the victim \cite{hernandez_that_2023}. In a similar incident, a finance worker in a multinational company was tricked into paying \$25 millions to fraudsters using deepfake technology to pose as the company’s chief financial officer in a video conference call \cite{deepfake_HongKong_scam}. 

Addressing the challenge of audio deepfakes, a number of audio spoof detection (ASD) methods were proposed to discriminate between bonafide and spoofed utterances. However, these ASD systems have been predominantly evaluated on ASVSpoof datasets (2015, 2017, 2019, 2021) \cite{wu2014asvspoof, delgado2018asvspoof, wang2020asvspoof, yamagishi2021asvspoof}. With the exception of the ASVSpoof 2021 dataset, these corpora have been curated within controlled settings which may not accurately depict real-world conditions. Recently, ASVspoof5 Challenge was started, and unlike previous ASVspoof databases, ASVspoof 5 database is built from crowd-sourced data collected in diverse acoustic conditions using Multilingual Librispeech (MLS) English partition. This database consists of 32 different spoofing attacks (A01-A32) and 11 codec and compression conditions (C01-C11). In addition to the use of new spoofing attacks implemented using the latest text-to- speech (TTS) synthesis and voice conversion (VC) algorithms, adversarial attacks are introduced for the first time and combined with spoofing attacks. For more detail, the reader is referred to Wang et al. \cite{Wang2024_ASVspoof5}.

This paper describes our submission to the ASVspoof Challenge and aims to investigate the performance of Audio Spoof Detection, trained using data augmentation through laundering attacks, on the ASVSpoof 5 database \cite{Wang2024_ASVspoof5}. For that purpose, we randomly selected 10\% of the audio files from the ASVspoof 5 train database (Non-Augmented data) \cite{Wang2024_ASVspoof5}. These audio files are then passed through a number of laundering attacks, including noise addition, reverberation, recompression, resampling, filtering, to generate Augmented ASVSpoof 5 train database (Augmented data). After that, we trained AASIST \cite{jung2022aasist} on Augmented data, and evaluated it on ASVSpoof 5 eval database by submitting the scores to the ASVspoof 5 Challenge \cite{Wang2024_ASVspoof5}. We hypothesize that the performance of AASIST will improve after training on Augmented data.

To benchmark AASIST on ASVSpoof 5, this paper presents the following contributions:

\begin{itemize}
    \item We trained AASIST on the Augmented ASVSpoof 5 train database (Augmented data), and evaluated it on the ASVSpoof 5 eval database.
    
    \item A detailed description of the performance of AASIST system, trained on Augmented data, is provided using 4 metrics namely, minDCF, actDCF, Cllr, and EER.

    \item A detailed breakdown of the results is provided in terms of Attacks vs Codecs.
\end{itemize}

A brief review of the relevant literature in the space of audio spoof detection (ASD) and the robustness of ASD systems is provided in Section \ref{related}. Section \ref{aasist} describes our selected method (AASIST) for this study. In Section \ref{data_augmention_laundering}, we discuss the process of generating augmentation data through laundering attacks. After that, the experimental setup and results are discussed in Sections \ref{exp_setup} and \ref{results}, respectively.

\begin{figure*}[!h]
\centering
\includegraphics[width=1\linewidth]{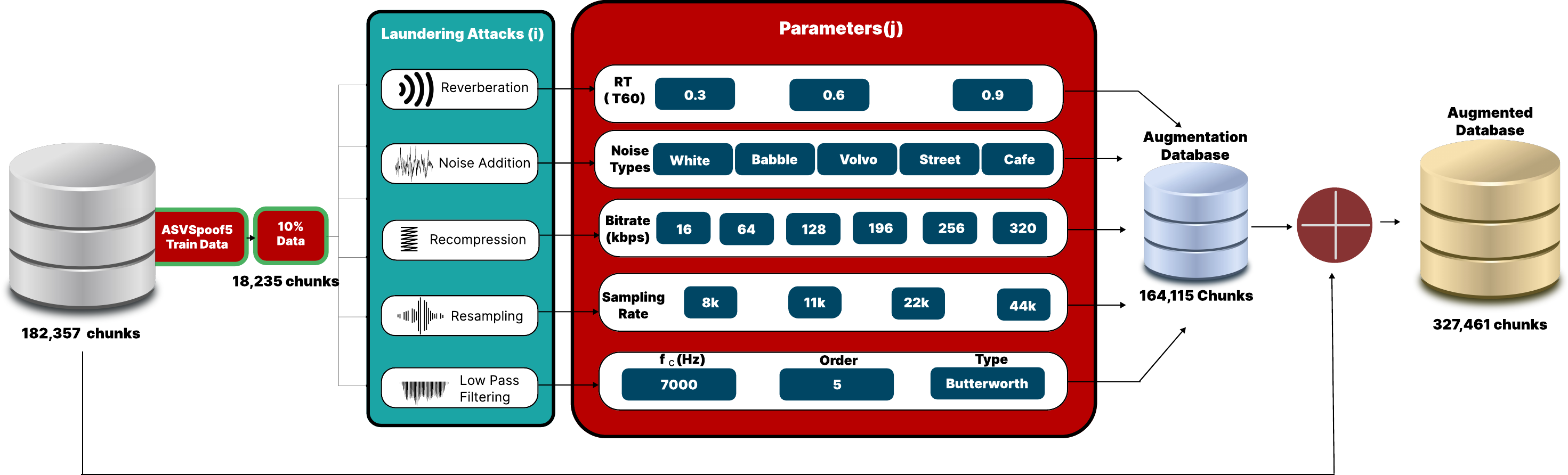}
\caption{Augmented Data Block Diagram: \normalfont ASVSpoof 5 train is the input database. First column describes the laundering attacks (i). Second column describes the parameters (j) for each laundering attack. Third column describes the generated Augmentation Database, which is then added to ASVspoof 5 train database to create the Augmented Database.}
\label{fig:augmented_data}
\end{figure*}

\section{Related Work} \label{related}
A significant amount of research has undergone to develop strategies that can detect audio spoofs reliably. These strategies can be broadly classified into three categories \cite{ali2023protecting, khan2023battling, balamurali2019toward, kamble2020advances}, 
\begin{enumerate}
    \item Conventional Machine Learning Approaches
    \item Representation Learning Approaches
    \item End-to-end Learning Approaches 
\end{enumerate}

Conventional machine learning (ML)-based approaches for audio spoof detection typically consist of two parts. The first part deals with hand-crafted feature extraction (front-end) and the second part consists of a model (back-end) that determines the authenticity of the audio signal \cite{balamurali2019toward, khan2023battling}. Examples of such systems include CQCC-GMM, LFCC-GMM \cite{yamagishi2021asvspoof, liu_asvspoof_2023} etc.

Representation learning approaches work either in the form of feature learning or as a pattern classifier. In representation learning, these methods use deep learning to generate a representation for the specific task and then use some classifier to discriminate between bonafide and spoof audios. Examples of feature learning include Qian et al. \cite{qian2016deep}. In pattern classification, hand-crafted features are extracted first and then a deep learning method is used as a classifier. Examples of pattern classification include LFCC-LCNN \cite{lavrentyeva2019stc}, OCSoftmax \cite{zhang2021one} etc. 

End-to-end learning approaches for audio spoof detection operate directly upon raw waveform inputs, streamlining the training and evaluation process. These methods use deep neural networks to learn a representation from raw audio input and then contain fully connected layers at the end for classification task. Examples of such systems include RawNet2 architecture \cite{jung2020improved, tak2021end}, RawGat-ST \cite{tak21_asvspoof}, AASIST \cite{jung2022aasist}. 

Several studies have explored the performance of audio spoof detection in acoustically degraded conditions and in the wild audio data. Muller \textit{et al.} \cite{muller_does_2022} re-implemented twelve popular architectures trained on ASVSpoof 2019 database and evaluated them on an in-the-wild database, consisting of audio data sourced from the internet. The authors demonstrated that the performance of ASD systems degrades by up to a thousand percent on such real-world data. However, Hashim \textit{et al.} \cite{ali2024audio} argued that the audio spoofs that are available online have undergone a number of post-processing steps, such as reverberation, recompression, additive noise, etc. As a result, an in-the-wild audio sourced from the internet could just be a clean audio file that has been subjected to laundering attacks. This led Hashim \textit{et al.} \cite{ali2024audio} to evaluate seven ASD systems on a laundered (noisy) database, called ``ASVSpoof Laundered Database''. The authors created this database by passing the audio files in ASVSpoof 2019 LA eval database through multiple laundering attacks. The authors demonstrated that the performance of all seven ASD systems degrade significantly in the presence of laundering attacks.

Considering the fact that (1) the ASVSpoof 5 database is crowd-sourced and consists of audio data collected in diverse acoustic conditions. (2) ASVSpoof 5 database contains audio file with varying codecs and compression conditions applied to them. We propose to train a baseline model from ASVSpoof 5 database on an Augmented data, generated by applying various laundering attacks to it (Section \ref{data_augmention_laundering}). We hypothesize that training AASIST system on an Augmented will improve its performance on ASVspoof 5 eval database. 

\section{AASIST System} \label{aasist}
AASIST \cite{jung2022aasist} is a baseline system in the ASVSpoof 5 Challenge. It used a RawNet2-based encoder \cite{jung2020improved, tak2021end} to extract spectro-temporal features from raw waveform inputs. First, the authors proposed a variant of the graph attention layer, known as the heterogeneous stacking graph attention layer'' (HS-GAL). This layer facilitates the concurrent modeling of heterogeneous (spectral and temporal) graph representations to create a single representation from them. HS-GAL comprises two
components, namely heterogeneous attention and a stack node. In heterogeneous attention, the authors use three different projection vectors to calculate the attention weights for the heterogeneous graph. After that, the stack node merges the information that spanned the relationship between the spectral and temporal domains. Additionally, the authors proposed a ``max graph operation'' (MGO), and a readout operation. Max graph operation (MGO) utilizes two parallel branches where the element-wise maximum is applied to the output of the two branches. This procedure aims to detect various artifacts introduced by spoofing in spoofed speech. Ultimately, CM output scores are generated using a readout operation and a hidden linear output layer comprising two class predictions: bonafide or spoof.

\section{Data Augmentation through Laundering Attacks} \label{data_augmention_laundering}
To improve the performance of audio spoof detection in real-world settings, we propose to train the AASIST \cite{jung2022aasist} system on a database augmented with laundering attacks. The idea is borrowed from Hashim \textit{et al.} \cite{ali2024audio} that an in-the-wild audio is just clean audio subjected to different types of laundering attacks, including noise addition, reverberation, and recompression, etc. For that purpose, 10\% of the audio files are randomly selected from the ASVSpoof 5 train database. This amounts to a total of 18235 audio files. Five different types of laundering attacks are then added to these audio files to create the augmentation data. First, reverberation noise is added with reverberation time (RT60) randomly chosen between 0.3s, 0.6s, and 0.9s. Second, the 10\% audio files are attacked with additive noise; babble noise, volvo noise, white noise, cafe noise, and street noise. Each noise is added to all the selected 10\% audio files with randomly chosen SNR levels between 0dB, 10dB, and 20dB, creating a total of 5 copies of the selected 10\% audio files. The third laundering attack that we added to the selected audio files is a recompression noise. The audio files in the ASVspoof 5 database are in FLAC format with a bitrate of 132 kbit/s. We first uncompressed the audio files from FLAC to WAV format. After that, the WAV audio files are compressed to MP3 format using bit rates randomly chosen between 16, 64, 128, 192, 256, and 320 kbit/s. Thereafter, all the audio files are uncompressed to WAV and compressed back to FLAC format. As a fourth laundering attack, we added resampling noise to the selected audio files. Taking into account the sampling rate of the original signal (16 KHz), the selected 10\% audio files from ASVspoof5 database were resampled with a sampling rate randomly chosen between 8000 Hz, 11025 Hz, 22050 Hz and 44100 Hz. Finally, the selected 10\% audio files are passed through a low pass butter-worth filter with a cut-off frequency of 8000 Hz and order 5. This process creates a total of 9 copies of the 10\% audio files randomly selected from the ASVSpoof 5 train database, five for additive noise laundering attacks and one for the remaining laundering attacks. This amounts to a total of 164,115 audio files in Augmentation data. This Augmentation data is then added to the ASVspoof 5 train database to create the Augmented database. The process of creating Augmented data is also illustrated in figure \ref{fig:augmented_data}.

\section{Experimental Setup} \label{exp_setup}
The goal of our experiments is to verify the hypothesis, mentioned in section \ref{intro}, that whether training the AASIST system using data augmentation through laundering attacks improve its performance.

\subsection{Training and Evaluation} \label{train_eval}
To verify our hypothesis, we trained the AASIST system on the augmented ASVSpoof 5 train database. It is developed by adding the augmentation data created in section \ref{data_augmention_laundering} to ASVSpoof 5 train database. We call this database an Augmented database. This database consists of a total of 327,461 audio files, of which 35,404 are bonafide and 311,068 are spoof. The detail of the Augmented ASVSpoof train database is given in Figure \ref{fig:augmented_data}. Furthermore, to avoid over-fitting, the ASVSpoof 5 development database is used as a validation. It is important to note here that the ASVSpoof 5 development database (validation set) does not contain any laundering attacks. The reason for this configuration is to achieve a more generalized performance while also achieving comparatively good results on the clean ASVSpoof 5 eval database. Once the AASIST system is trained, it is evaluated on ASVSpoof 5 eval database by submitting the scores to the ASVspoof 5 Challenge.

\subsection{Evaluation Metric}
ASVspoof 5 Challenge uses four metrics for evaluation, namely minimum Detection Cost Function (minDCF), actual Detection Cost Function (actDCF), cost of log-likelihood ratios ($C_{llr}$), and equal error rate (EER). The details of each evaluation metric are given in the ASVspoof 5 summary paper \cite{Wang2024_ASVspoof5}.

\section{Results} \label{results}
This section discusses the findings of our experiments. As mentioned in section \ref{exp_setup}, the goal of our experiments is to study the performance of the AASIST system on the ASVSpoof 5 Challenge database, when it is trained using data augmentation through laundering attacks. For that purpose, a detailed breakdown of the results is provided in terms of spoofing attacks (A17-A32) vs codecs and compression conditions (C00-C11) in terms of minDCF, actDCF, $C_{llr}$, and EER in tables \ref{table:detailed_mindcf}, \ref{table:detailed_actdcf} \ref{table:detailed_cllr}, and \ref{table:detailed_eer}. 

\begin{table}
\centering
\caption{Performance of AASIST on ASVSpoof5 database in terms of pooled minDCF, actDCF, $C_{llr}$, EER}

    \begin{tabular}{| c | c | c | c | c |}
     \hline
     {} & minDCF & actDCF & $C_{llr}$ & EER\\
      \hline
     AASIST & 0.662 & 0.931 & 2.486 & 25.319\\
     \hline
     
    \end{tabular}

\label{table:aasist_performance} 
\end{table}

Table \ref{table:aasist_performance} shows the results of AASIST, trained using data augmentation through laundering attacks, in terms of pooled minDCF, actDCF, $C_{llr}$, and EER. The table shows that AASIST achieves minDCF value of 0.662, actDCF value of 0.931, $C_{llr}$ value of 2.486, and EER value of 25.319\%.

\begin{table}
\centering
\caption{Performance of AASIST on individual spoofing attacks (A17-A32) in terms of pooled minDCF, actDCF, $C_{llr}$, EER. Bold entries show the top-5 worst performances.}

    \begin{tabular}{| c | c | c | c | c |}
     \hline
     Spoofing Attack & minDCF & actDCF & $C_{llr}$ & EER\\
     \hline
     A17 & 0.428 & 0.963 & 1.846 & 14.944\\
     \hline
     A18 & \textbf{0.865} & \textbf{0.998} & \textbf{3.098} & \textbf{30.232}\\
     \hline
     A19 & \textbf{1.0} & \textbf{1.0} & \textbf{4.650} & \textbf{56.669}\\
     \hline
     A20 & \textbf{0.994} & \textbf{1.0} & \textbf{3.844} & \textbf{43.885}\\
     \hline
     A21 & 0.346 & 0.879 & 1.461 & 12.296\\
     \hline
     A22 & 0.357 & 0.924 & 1.556 & 12.661\\
     \hline
     A23 & 0.481 & 0.951 & 2.0 & 16.588\\
     \hline
     A24 & 0.268 & 0.762 & 1.143 & 9.905\\
     \hline
     A25 & 0.711 & 0.994 & 2.719 & 24.818\\
     \hline
     A26 & \textbf{0.857} & \textbf{0.999} & \textbf{3.130} & \textbf{29.960}\\
     \hline
     A27 & 0.667 & 0.992 & 2.591 & 23.830\\
     \hline
     A28 & 0.626 & \textbf{0.998} & 2.544 & 21.678\\
     \hline
     A29 & 0.173 & 0.465 & 0.641 & 6.687\\
     \hline
     A30 & \textbf{1.0} & \textbf{1.0} & \textbf{3.825} & \textbf{41.245}\\
     \hline
     A31 & 0.547 & 0.986 & 2.205 & 19.307\\
     \hline
     A32 & 0.766 & 0.994 & 2.868 & 27.679\\
     \hline

    \end{tabular}

\label{table:aasist_performance_attacks} 
\end{table}

Table \ref{table:aasist_performance_attacks} shows the results of the modified AASIST system on individual attacks in terms of pooled minDCF, actDCF, $C_{llr}$ and EER. For each metric, the bold entries depict the top-5 worst performances among spoofing attacks A17-A32. Our observations can be summarized as follows.

\begin{itemize}
    \item AASIST system shows worst performance on A18, A19, A20, A26, and A30 spoofing attacks with minDCF values of 0.865, 1.0, 0.994, 0.857, and 1.0 respectively.

    \item A18 consists of A17 attack plus Malafide adversarial attack \cite{panariello2023malafide}, whereas, A20 consists of A12 + Malafide adversarial attack \cite{panariello2023malafide}. We can observe that the modified AASIST system shows good performance on A17 attack with minDCF value of 0.428, however, addition of Malafide adversarial attack degrades the performance for both A18 and A20 spoofing attacks.
    
    \item A30 attack is a combination of A18 attack and Malacopula adversarial attack \cite{Wang2024_ASVspoof5}. In other words, it is a combination of A17, Malafide and Malacopula attacks. The modified AASIST system achieves a minDCF value of 1.0 on this attack. We can see a gradual degradation in performance with the addition of Malafide and Malacopula attacks, from 0.428 on A17 to 0.865 on A17 + Malafide to 1.0 on A17 + Malafide + Malacopula.

    \item The modified AASIST system does not show good performance on A26 attack (minDCF equal to 0.857), which is a combination of A16 attack and background noise. It is surprising for a system trained on augmented data with background noise. 

    \item The modified AASIST system also shows one of the worst performance on A19 attack (minDCF equal to 1), which is a TTS attack based on MaryTTS \cite{steiner2017creating}. A19 is the only attack in top-5 worst attacks that does not have any adversarial attack or background noise added to it. 

    \item Moreover, actDCF, $C_{llr}$ and EER also show worst performances on the same attacks i.e., A18, A19, A20, A26, and A30. Furthermore, actDCF values are close to or equal to 1 (the worst case value) for most of the attacks, except A24 and A29. This suggests that AASIST's outputs are either larger or smaller than the decision threshold decided by the priors and decision costs.

    \item The modified AASIST system performs the best on A29 attack with a minDCF value of 0.173. A29 is a TTS attack using pre-trained XTTS \cite{casanova2024xtts}.
    
\end{itemize}

\begin{table}
\centering
\caption{Performance of AASIST under different codec and compression conditions in terms of pooled minDCF, actDCF, $C_{llr}$, EER. Bold entries show the top-5 worst performances.}

    \begin{tabular}{| c | c | c | c | c |}
     \hline
     Codec & minDCF & actDCF & $C_{llr}$ & EER\\
     \hline
     C00 & 0.383 & 0.902 & \textbf{2.616} & 16.922\\
     \hline
     C01 & 0.536 & 0.955 & \textbf{2.668} & 21.639\\
     \hline
     C02 & 0.535 & \textbf{0.966} & 2.584 & 22.426\\
     \hline
     C03 & 0.533 & \textbf{0.981} & \textbf{3.136} & 21.946\\
     \hline
     C04 & \textbf{0.627} & \textbf{0.984} & \textbf{3.155} & \textbf{27.110}\\
     \hline
     C05 & 0.402 & 0.896 & 2.420 & 18.023\\
     \hline
     C06 & 0.573 & 0.913 & 2.436 & 21.211 \\
     \hline
     C07 & \textbf{0.637} & \textbf{0.984} & \textbf{3.061} & \textbf{27.680}\\
     \hline
     C08 & \textbf{0.705} & 0.913 & 2.066 & \textbf{32.466}\\
     \hline
     C09 & \textbf{0.693} & \textbf{0.968} & 1.923 & \textbf{28.753}\\
     \hline
     C10 & \textbf{0.711} & 0.913 & 1.583 & \textbf{29.321}\\
     \hline
     C11 & 0.550 & 0.864 & 1.734 & 23.701\\
     \hline

    \end{tabular}

\label{table:aasist_performance_codecs} 
\end{table}

Table \ref{table:aasist_performance_codecs} displays the performance of the modified AASIST system (trained on Augmented database) under different codec and compression conditions in terms of pooled minDCF, actDCF, $C_{llr}$ and EER. Again, for each metric, the bold entries depict the top-5 worst performances among codec and compression conditions (C00-C11). Our observations can be summarized as follows.

\begin{itemize}
    \item AASIST system shows worst performance under C04, C07, C08, C09, and C10 with minDCF values of 0.627, 0.637, 0.705, 0.693, and 0.711 respectively.

    \item C08, C09, and C10 has a bandwidth of 8 kHz. This suggests that the modified AASIST system does not perform good when the sampling rate is 8 kHz.

    \item Moreover, C04 uses Ecodec \cite{defossez2022high} codec and C07 uses Encodec \cite{defossez2022high} + mp3 codec. AASIST achieves minDCF value of 0.625 on C04 and 0.637 on C07 respectively.

    \item The modified AASIST system performs the best under no codec and compression condition (C00) and mp3 codec (C05) with a bitrate range of 45-256. The reason for good performance in C04 is that the Augmented data (Section \ref{data_augmention_laundering}) contains recompression laundering attack with various bitrates. 
\end{itemize}

The codebase for generating augmented data and training and evaluation of modified AASIST can be found at the following GitHub repositories {\footnote{https://github.com/hashim19/Rob-ASD}} {\footnote{https://github.com/suryasubbu/Audio-Laundering-Attacks}}.
  

\begin{table*}
\centering
\caption{Detailed Result Break Down for AASIST in terms of Minimum Detection Cost Function (min DCF): Attacks vs Codecs}

\begin{tabular}{ccccccccccccc}
\toprule
       & - & codec-1 & codec-10 & codec-11 & codec-2 & codec-3 & codec-4 & codec-5 & codec-6 & codec-7 & codec-8 & codec-9 \\ 
\midrule
 A17   &  0.101 &  0.280 &  0.509 &  0.257 &  0.271 &  0.339 &  0.320 &  0.112 &  0.326 &  0.368 &  0.495 &  0.548\\ 
 A18   &  0.492 &  0.751 &  0.936 &  0.622 &  0.710 &  0.708 &  0.803 &  0.556 &  0.696 &  0.812 &  0.796 &  0.780\\ 
 A19   &  1.0 &  1.0 &  1.0 &  1.0 &  1.0 &  1.0 &  1.0 &  1.0 &  1.0 &  1.0 &  1.0 &  1.0\\ 
 A20   &  0.999 &  0.961 &  0.961 &  1.0 &  0.997 &  0.972 &  1.0 &  1.0 &  0.967 &  1.0 &  1.0 &  0.885\\ 
 A21   &  0.044 &  0.180 &  0.264 &  0.148 &  0.198 &  0.289 &  0.327 &  0.052 &  0.239 &  0.324 &  0.342 &  0.290\\ 
 A22   &  0.076 &  0.208 &  0.312 &  0.166 &  0.220 &  0.254 &  0.312 &  0.089 &  0.254 &  0.313 &  0.373 &  0.359\\ 
 A23   &  0.125 &  0.275 &  0.337 &  0.226 &  0.281 &  0.320 &  0.415 &  0.150 &  0.325 &  0.428 &  0.366 &  0.277\\ 
 A24   &  0.030 &  0.168 &  0.449 &  0.165 &  0.162 &  0.218 &  0.173 &  0.050 &  0.168 &  0.193 &  0.435 &  0.552\\ 
 A25   &  0.222 &  0.638 &  0.971 &  0.545 &  0.662 &  0.674 &  0.748 &  0.263 &  0.491 &  0.762 &  0.927 &  1.0\\ 
 A26   & 0.429 &  0.723 &  0.981 &  0.758 &  0.754 &  0.748 &  0.859 &  0.458 &  0.688 &  0.852 &  0.934 &  0.988\\ 
 A27   &  0.328 &  0.490 &  0.845 &  0.625 &  0.488 &  0.397 &  0.653 &  0.362 &  0.539 &  0.691 &  0.801 &  0.755\\ 
 A28   &  0.234 &  0.451 &  0.790 &  0.503 &  0.564 &  0.492 &  0.513 &  0.246 &  0.508 &  0.536 &  0.663 &  0.751\\ 
 A29   &  0.006 &  0.104 &  0.241 &  0.079 &  0.066 &  0.112 &  0.110 &  0.007 &  0.063 &  0.144 &  0.322 &  0.305\\ 
 A30   &  0.838 &  0.969 &  1.0 &  0.999 &  0.931 &  0.907 &  0.999 &  0.895 &  0.989 &  0.999 &  1.0 &  1.0\\ 
 A31   &  0.236 &  0.326 &  0.539 &  0.498 &  0.262 &  0.255 &  0.509 &  0.289 &  0.462 &  0.562 &  0.612 &  0.450\\ 
 A32   &  0.422 &  0.629 &  0.907 &  0.694 &  0.590 &  0.496 &  0.820 &  0.454 &  0.622 &  0.828 &  0.915 &  0.831\\ 
\bottomrule
\end{tabular}

\label{table:detailed_mindcf} 
\vspace{-1em}
\end{table*}

\begin{table*}
\centering
\caption{Detailed Result Break Down for AASIST in terms of Actual Detection Cost Function (act DCF): Attacks vs Codecs}

\begin{tabular}{cccccccccccccc}
\toprule
       & - & codec-1 & codec-10 & codec-11 & codec-2 & codec-3 & codec-4 & codec-5 & codec-6 & codec-7 & codec-8 & codec-9 \\ 
\midrule
 A17   &  0.962 &  0.989 &  0.935 &  0.893 &  0.988 &  0.996 &  0.996 &  0.936 &  0.955 &  0.992 &  0.923 &  0.981\\ 
 A18   &  1.0 &  1.0 &  0.997 &  0.998 &  1.0 &  1.0 &  1.0 &  1.0 &  0.995 &  1.0 &  0.996 &  0.996\\ 
 A19   &  1.0 &  1.0 &  1.001 &  1.0 &  1.0 &  1.0 &  1.0 &  1.0 &  1.0 &  1.0 &  1.0 &  1.0\\ 
 A20   &  1.0 &  1.0 &  1.0 &  1.0 &  1.0 &  1.0 &  1.0 &  1.0 &  0.999 &  1.0 &  1.0 &  1.0\\ 
 A21   &  0.920 &  0.970 &  0.673 &  0.588 &  0.974 &  0.995 &  0.996 &  0.855 &  0.897 &  0.995 &  0.711 &  0.861\\ 
 A22   &  0.951 &  0.972 &  0.800 &  0.720 &  0.966 &  0.994 &  0.993 &  0.922 &  0.950 &  0.994 &  0.809 &  0.937\\ 
 A23   &  0.998 &  0.995 &  0.834 &  0.817 &  0.997 &  1.0 &  1.0 &  0.996 &  0.910 &  0.999 &  0.829 &  0.903\\ 
 A24   &  0.551 &  0.834 &  0.881 &  0.684 &  0.890 &  0.934 &  0.945 &  0.564 &  0.685 &  0.929 &  0.836 &  0.977\\ 
 A25   &  0.991 &  0.999 &  1.0 &  0.964 &  1.0 &  1.0 &  1.0 &  0.990 &  0.982 &  1.0 &  0.999 &  1.0\\ 
 A26   &  0.999 &  0.999 &  0.998 &  0.998 &  0.999 &  1.0 &  1.0 &  0.998 &  0.998 &  1.0 &  1.0 &  1.0\\ 
 A27   &  0.990 &  0.994 &  0.983 &  0.988 &  0.997 &  0.996 &  0.999 &  0.987 &  0.986 &  0.999 &  0.984 &  0.994\\ 
 A28   &  0.999 &  0.999 &  0.993 &  0.989 &  0.999 &  1.0 &  0.999 &  0.997 &  0.997 &  0.999 &  0.994 &  0.999\\ 
 A29   &  0.106 &  0.553 &  0.591 &  0.264 &  0.673 &  0.798 &  0.831 &  0.143 &  0.303 &  0.841 &  0.587 &  0.854\\ 
 A30   &  0.999 &  1.0 &  1.0 &  1.0 &  1.0 &  1.0 &  1.0 &  1.0 &  0.999 &  1.0 &  1.0 &  1.0\\ 
 A31   &  0.992 &  0.987 &  0.948 &  0.969 &  0.991 &  0.996 &  1.0 &  0.990 &  0.986 &  0.999 &  0.965 &  0.983\\ 
 A32   &  0.991 &  0.996 &  0.994 &  0.987 &  0.996 &  0.998 &  0.999 &  0.990 &  0.988 &  0.998 &  0.998 &  0.998\\ 
\bottomrule
\end{tabular}

\label{table:detailed_actdcf} 
\vspace{-1em}
\end{table*}

\begin{table*}
\centering
\caption{Detailed Result Break Down for AASIST in terms of Cost of Log-Likelihood Ratio ($C_{llr}$): Attacks vs Codecs}

\begin{tabular}{cccccccccccccc}
\toprule
       & - & codec-1 & codec-10 & codec-11 & codec-2 & codec-3 & codec-4 & codec-5 & codec-6 & codec-7 & codec-8 & codec-9 \\ 
\midrule 
 A17   &  1.765 &  2.087 &  1.196 &  1.086 &  2.023 &  2.741 &  2.453 &  1.551 &  1.758 &  2.413 &  1.584 &  1.655\\ 
 A18   &  3.545 &  3.390 &  1.937 &  2.073 &  3.192 &  3.754 &  3.661 &  3.330 &  3.106 &  3.516 &  2.259 &  2.085\\ 
 A19   &  5.575 &  4.944 &  2.791 &  3.706 &  4.640 &  5.110 &  4.870 &  5.295 &  5.049 &  4.689 &  3.392 &  3.164\\ 
 A20   &  4.633 &  3.925 &  2.014 &  2.931 &  3.800 &  4.299 &  4.442 &  4.372 &  4.021 &  4.302 &  2.801 &  2.322\\ 
 A21   &  1.275 &  1.700 &  0.675 &  0.635 &  1.721 &  2.581 &  2.487 &  1.071 &  1.333 &  2.349 &  1.065 &  1.011\\ 
 A22   &  1.528 &  1.701 &  0.809 &  0.756 &  1.683 &  2.402 &  2.364 &  1.310 &  1.493 &  2.223 &  1.191 &  1.201\\ 
 A23   &  2.292 &  2.145 &  0.857 &  0.960 &  2.143 &  2.801 &  2.859 &  2.088 &  1.849 &  2.748 &  1.285 &  1.038\\ 
 A24   &  0.636 &  1.340 &  1.077 &  0.726 &  1.353 &  1.948 &  1.673 &  0.682 &  0.895 &  1.642 &  1.437 &  1.644\\ 
 A25   &  2.488 &  3.193 &  2.032 &  1.789 &  3.103 &  3.704 &  3.521 &  2.341 &  2.414 &  3.375 &  2.574 &  2.622\\ 
 A26   &  3.295 &  3.367 &  2.142 &  2.360 &  3.280 &  3.864 &  3.785 &  3.001 &  3.139 &  3.602 &  2.594 &  2.611\\ 
 A27   &  2.746 &  2.650 &  1.780 &  2.011 &  2.622 &  2.877 &  3.303 &  2.467 &  2.492 &  3.224 &  2.375 &  2.027\\ 
 A28   &  2.686 &  2.716 &  1.696 &  1.791 &  2.885 &  3.304 &  3.065 &  2.367 &  2.604 &  2.944 &  1.975 &  2.053\\ 
 A29   &  0.135 &  0.760 &  0.597 &  0.312 &  0.762 &  1.352 &  1.219 &  0.159 &  0.335 &  1.340 &  1.023 &  1.021\\ 
 A30   &  4.258 &  3.928 &  2.739 &  3.044 &  3.655 &  4.167 &  4.433 &  4.057 &  3.880 &  4.368 &  3.237 &  2.863\\ 
 A31   &  2.463 &  2.152 &  1.248 &  1.696 &  1.953 &  2.425 &  2.940 &  2.323 &  2.267 &  2.922 &  1.869 &  1.426\\ 
 A32   &  3.037 &  3.035 &  1.913 &  2.185 &  2.842 &  3.142 &  3.720 &  2.801 &  2.753 &  3.615 &  2.622 &  2.193\\ 
\bottomrule
\end{tabular}

\label{table:detailed_cllr} 
\end{table*}

\begin{table*}
\centering
\caption{Detailed Result Break Down for AASIST in terms of Equal Error Rate (EER, \%): Attacks vs Codecs}

\begin{tabular}{cccccccccccccc}
\toprule
       & - & codec-1 & codec-10 & codec-11 & codec-2 & codec-3 & codec-4 & codec-5 & codec-6 & codec-7 & codec-8 & codec-9 \\ 
\midrule
 A17   &   3.565  &   9.819  &  18.515  &   9.628  &   9.924  &  12.347  &  11.644  &   3.952  &  11.410  &  13.483  &  21.895  &  20.135 \\ 
 A18   &  17.284  &  26.411  &  34.084  &  22.400  &  26.713  &  26.455  &  30.700  &  19.609  &  24.211  &  30.614  &  31.124  &  29.722 \\ 
 A19   &  61.312  &  59.229  &  53.364  &  56.691  &  58.513  &  57.177  &  58.527  &  61.932  &  60.287  &  58.056  &  54.248  &  53.555 \\ 
 A20   &  37.779  &  37.947  &  35.531  &  39.722  &  38.810  &  37.321  &  48.046  &  40.072  &  40.242  &  48.895  &  41.121  &  34.015 \\ 
 A21   &   1.573  &   6.397  &   9.604  &   6.427  &   6.984  &  10.353  &  11.533  &   1.960  &   8.378  &  11.674  &  17.415  &  10.582 \\ 
 A22   &   2.637  &   7.518  &  11.259  &   6.613  &   8.057  &   9.393  &  11.362  &   3.303  &   8.896  &  11.539  &  18.152  &  13.254 \\ 
 A23   &   4.358  &   9.555  &  12.255  &   8.976  &   9.753  &  11.526  &  15.083  &   5.331  &  11.517  &  15.796  &  19.362  &  10.287 \\ 
 A24   &   1.122  &   6.417  &  17.083  &   6.511  &   6.184  &   8.825  &   6.140  &   1.783  &   6.101  &   7.324  &  21.826  &  20.472 \\ 
 A25   &   7.991  &  22.438  &  35.915  &  21.066  &  24.207  &  24.750  &  28.772  &   9.542  &  17.077  &  29.210  &  37.717  &  40.695 \\ 
 A26   &  15.500  &  25.830  &  39.099  &  29.518  &  28.642  &  28.529  &  33.128  &  16.901  &  23.817  &  32.319  &  37.709  &  40.906 \\ 
 A27   &  12.285  &  17.944  &  31.584  &  24.525  &  17.970  &  14.797  &  25.854  &  13.434  &  19.016  &  27.701  &  33.501  &  28.704 \\ 
 A28   &   8.265  &  16.136  &  28.626  &  18.518  &  20.560  &  18.188  &  18.741  &   9.065  &  17.591  &  19.371  &  26.164  &  28.198 \\ 
 A29   &   0.291  &   4.103  &   8.539  &   3.394  &   2.504  &   4.122  &   4.185  &   0.337  &   2.476  &   5.424  &  19.015  &  11.500 \\ 
 A30   &  31.455  &  37.730  &  52.931  &  42.390  &  37.175  &  35.242  &  48.279  &  34.556  &  36.038  &  50.534  &  50.695  &  46.714 \\ 
 A31   &   8.563  &  11.977  &  19.686  &  19.176  &  10.144  &   9.505  &  19.258  &  10.655  &  16.064  &  22.091  &  25.826  &  16.418 \\ 
 A32   &  16.101  &  23.406  &  34.575  &  27.430  &  22.673  &  18.744  &  34.740  &  17.714  &  21.817  &  35.607  &  38.642  &  31.662 \\ 
\bottomrule
\end{tabular}

\label{table:detailed_eer} 
\end{table*}

\section{Conclusion}
We trained a baseline model, AASIST, on an augmented database as our submission to the ASVspoof 5 Challenge. This database is created by randomly selecting 10\% of the audio files from the ASVspoof 5 train database, and applying different laundering attacks, including reverberation, noise addition, recompression, resampling, and low pass filtering, to generate an Augmented database. We achieved a pooled minDCF value of 0.662 and an EER of 25.319\% in the ASVspoof 5 challenge. In addition, we studied the results of our system on individual spoofing attacks. We observed that the system shows the worst performance in the presence of adversarial attacks in the audio files, with minDCF values of 0.865 for A18 attack, 0.994 for A30 attack, and 1.0 for A30 attack. Furthermore, we also studied the performance of our system under different codec and compression conditions. We observed that our system shows the worst performance when the sampling rate is 8 kHz (C08, C09, C10) and in the presence of Encodec codec (C04, C07) with a minDCF value of 0.627 for C04, 0.637 for C07, 0.705 for C08, 0.693 for C09, and 0.711 for C10 respectively.

\bibliographystyle{IEEEbib}
\bibliography{references}

%

\end{document}